\documentclass[12pt]{article}

\usepackage[numbers,sort&compress]{natbib}
\usepackage{amsmath}
\usepackage{amssymb}
\usepackage{hyperref}
\usepackage[bf]{caption}
\usepackage{color}

\usepackage{sectsty}
\sectionfont{\large}
\subsectionfont{\normalsize}

\usepackage{float}

\newcommand{\ba}{\begin{aligned}}
\newcommand{\ea}{\end{aligned}}
\newcommand{\beq}{\begin{equation}}
\newcommand{\eeq}{\end{equation}}
\def\ben{\begin{equation*}}
\def\een{\end{equation*}}
\newcommand{\beqs}{\begin{eqnarray}}
\newcommand{\eeqs}{\end{eqnarray}}

\newcommand{\sdot}{\hspace{-3pt}\cdot\hspace{-3pt}}

\def\be{\begin{equation}}
\def\ee{\end{equation}}
\def\bea{\begin{eqnarray}}
\def\eea{\end{eqnarray}}
\def\bsp{\be\begin{split}}

\def\a{\alpha}
\def\b{\beta}

\def\d{\delta}
\def\e{\epsilon}

\def\s{\sigma}

\makeatletter

\newcommand{\Rmnum}[1]{\expandafter\@slowromancap\romannumeral #1@}
\makeatother

\renewcommand{\title}[1]{\vbox{\center\LARGE{#1}}\vspace{5mm}}
\renewcommand{\author}[1]{\vbox{\center\large{#1}}\vspace{5mm}}

\everymath{\displaystyle}

\begin{document}

\begin{titlepage}
\begin{flushright}
\vspace{10pt} \hfill{NCTS-TH/1505 }\vspace{20mm}
\end{flushright}
\begin{center}

{\Large \bf Comments on the evaluation of massless scattering}

\vspace{45pt}

{
\textbf{Carlos Cardona},$^a$ 
\textbf{Chrysostomos Kalousios},$^b$
\footnote[1]{
\href{mailto:carlosandres@mx.nthu.edu.tw }{\tt{carlosandres@mx.nthu.edu.tw}}\,,  
\href{mailto:ckalousi@ift.unesp.br}{\tt{ckalousi@ift.unesp.br}}
}
}
\\[15mm]

{\it\ 
${}^a\,$Physics Division, National Center for Theoretical Sciences, National Tsing-Hua University,
Hsinchu, Taiwan 30013, Republic of China.

\vspace{20pt}

${}^b\,$ICTP South American Institute for Fundamental Research\\
Instituto de F\'\i sica Te\'orica, UNESP-Universidade Estadual Paulista\\
R. Dr. Bento T. Ferraz 271 - Bl. II, 01140-070, S\~ao Paulo, SP, Brasil}\\
\vspace{10pt}

\vspace{20pt}

\end{center}

\vspace{40pt}

\centerline{{\bf{Abstract}}}
\vspace*{5mm}
\noindent
The goal of this work is threefold.  First, we give an expression of the most 
general five point integral on ${\cal M}_{0,n}$ in terms of Chebyshev 
polynomials.  Second, we choose a special kinematics that transforms the 
polynomial form of the scattering equations to a linear system of symmetric 
polynomials.  We then explain how this can be used to explicitly evaluate
arbitrary point integrals on ${\cal M}_{0,n}$.  Third, we comment on the recently presented method of companion matrices and we show its equivalence to the elimination theory and an algorithm previously developed by one of the authors.
\vspace{15pt}
\end{titlepage}

\newpage

\section{Introduction}

The last couple of years substantial progress has been made in understanding and calculating massless scattering of several field theories following the elegant work of Cachazo-He-Yuan (CHY)\cite{Cachazo:2013gna,Cachazo:2013hca,Cachazo:2013iea,Cachazo:2014nsa,Cachazo:2014xea}.  The CHY construction was proven for scalar $\phi^3$ and pure Yang-Mills by Dolan and Goddard in \cite{Dolan:2013isa}.  According to CHY, massless scattering at tree level in arbitrary dimensions can in general be described by the contour integral
\be \label{An=}
\mathcal{A}_n = \int \frac{{\rm d}^n \s}{{\rm vol}\, SL(2,\mathbb{C})}
\s_{ij}\s_{jk}\s_{ki}\hspace{-6pt} \prod_{a \neq i,j,k}\hspace{-6pt} \d(f_a) \hspace{2pt} I_n(k,\e,\s),
\ee
where $\s_i$ are complex variables, $\s_{ij}=\s_i-\s_j$, $k$ is the external momentum, $\e$ the helicities and $I_n$ depends on the theory. The index $n$ labels the external fields.  The appearance of the delta function in \eqref{An=} completely localizes the integral over the solutions of the so-called scattering equations
\be \label{scattering eqns}
f_a \equiv \sum_{b\neq a}^n \frac{s_{ab}}{\s_{ab}} = 0,
\ee
where $s_{ab}=(k_a+k_b)^2 = 2 k_a \sdot k_b$ are the kinematic invariants.  Examples and details about specific theories can be found in the original literature.  The scattering equations have initially appeared in the literature in the work of Fairlie and Roberts \cite{Fairlie,Roberts,Fairlie:2008dg} and then in the work of Gross and Mende \cite{Gross:1987ar}.

It has been by now well established that in order to evaluate scattering amplitudes in massless theories satisfying the scattering equations of CHY, there is no need to explicitly solve those equations. A number of interesting methods have been considered lately. The first of these approaches was given by one of the authors in \cite{Kalousios:2015fya} by using the polynomial form of the scattering equations found in \cite{Dolan:2014ega} and using the Vieta formulas to compute the sums of the roots of the polynomials in terms of its coefficients. In \cite{Cachazo:2015nwa} another interesting method was developed, in which any amplitude can be decomposed in terms of some well know expressions for $\phi^3$ theory. A different approach was considered in \cite{Baadsgaard:2015voa,Baadsgaard:2015ifa}, where based on some results from \cite{Bjerrum-Bohr:2014qwa} a direct matching between Feynman diagrams and integration measures in the scattering equation formalism of CHY was found.  Very recently a nice method has been developed in \cite{Huang:2015yka}, where the authors use the Gr\"{o}bner basis associated to the polynomial form of the scattering equations and through this new basis they constructed the so-called companion matrices which give the scattering amplitude in terms of traces over products of those matrices. \footnote{After the completion of this paper, a related approach appeared in \cite{Sogaard:2015dba} where the authors use the B\'ezoutian matrix to compute the amplitudes.}

In this note we employ three different ideas related to the scattering equations and contour integrals for massless scattering in arbitrary dimensions.  The first idea is a simple and interesting rewriting of the most general integral in the $n=5$ case.  This integral that depends on five cross ratios was previously evaluated in \cite{Kalousios:2015fya} through the construction of a generating function that gives integrals raised to different powers of the cross ratios.  Here we will give an alternative expression of the aforementioned fundamental quantity purely in terms of Chebyshev polynomials, by employing properties of the polynomials and making use of the fact that for $n=5$ the scattering equations have two solutions.

The second idea we will discuss is for general $n$ and special kinematics.  Studying amplitudes at special kinematics has previously appeared for example in \cite{Cachazo:2013iea,Dolan:2014ega,Kalousios:2013eca,Weinzierl:2014vwa,Lam:2014tga} and in some of the cases it has allowed for the explicit evaluation of the amplitudes \cite{Cachazo:2013iea,Kalousios:2013eca}.  Studying amplitudes at special kinematics is definitely interesting as it reveals features of the scattering equations and the amplitudes that are hidden in the complexity of the general solutions.  As a recent example of applications we will mention the interesting work of \cite{Dvali:2014ila}.  Motivated by the above, we will consider a special kinematics that will allow the linearization of the polynomial in nature scattering equations.  Then, any amplitude can be straightforwardly evaluated without the need to know the explicit solutions of the scattering equations.  An interesting interpretation of the number of solutions of the scattering equations, namely $(n-3)!$, will be also given.  It turns out that the special kinematics is the one previously considered in \cite{Kalousios:2013eca}.  Here we will generalize the results of \cite{Kalousios:2013eca} and we will provide a framework to the proof of previously numerically found expressions for the amplitudes.

The third idea we will discuss is related to the aforementioned method of companion matrices.  We will provide several comments and argue that the method is equivalent to the elimination theory of \cite{Dolan:2014ega} (that can be alternatively used to construct the companion matrices) and an algorithm presented by one of the authors in \cite{Kalousios:2015fya}.  The basic idea is that because of the explicit form of the companion matrices, evaluating any given amplitude by taking traces of companion matrices is essentially a neat way to isolate the appropriate coefficients of polynomials that the scattering equations satisfy.  The latter was also the observation in \cite{Kalousios:2015fya}, where the coefficients of polynomials where put in use through the well-known in mathematics Vieta formulas.

Each of the three ideas will occupy a separate section, whereas conclusions will be presented at the end.

\section{Five point integrals for general kinematics as Chebyshev polynomials}
In \cite{Kalousios:2015fya} the most general quantity, $P$, for five point amplitudes consistent with $SL(2,\mathbb{C})$ invariance was given in terms of five cross ratios
\be \label{fundamental}
P = \sum_{\rm{solutions}} \frac{1}{\rm{det}' \Phi}
\prod_{i=1}^5 \frac{1}{\s_{i,i+1}^2}
\left( \frac{\s_{i,i+2} \s_{i+1,i+4}}{\s_{i,i+1}\s_{i+2,i+4}} \right)^{\a_i},
\ee
where $\a_i$ are integers that can also be negative.  Here ${\rm{det}'} \Phi$ is the determinant of the matrix $\Phi_{ab} = \partial f_a / \partial \s_b$ after the removal of rows $j,k,l$ and columns $p,q,r$, divided by the quantity $(\s_{jk}\s_{kl}\s_{lj})(\s_{pq}\s_{qr}\s_{rp})$ as required by the $SL(2,\mathbb{C})$ invariance.  The sum runs over the solutions of the scattering equations \eqref{scattering eqns}.

In \cite{Kalousios:2015fya} it was shown that the expression \eqref{fundamental} can be neatly organized in terms of a generating function.  Here we find it interesting to give another expression of \eqref{fundamental} in terms of Chebyshev polynomials of the first and second kind, $T_n(x)$ and $U_n(x)$.  We find
\be \label{fundamental2}
P = \prod_{j=0}^5 f_j + {\sum}^{'} f_j f_k f_l f_m g_n g_p +{\sum}^{'} f_j f_k g_l g_m g_n g_p +{\sum}^{'} g_j g_k g_l g_m g_n g_p,
\ee
where by ${\sum}^{'}$ we denote that the sum runs over all values of indices from 0 to 5, such that no index appears more than once and all terms are different to each other.  Explicitly, the first sum means that $j<k<l<m,~n<p,~j\neq k \neq l \neq m \neq n \neq p$, the second sum means $j<k,~l<m<n<p,~j\neq k \neq l \neq m \neq n \neq p$ and the last sum means $j<k<l<m<n<p$.  All together there are 32 terms.  The functions appearing in \eqref{fundamental2} are given in terms of Chebyshev polynomials
\be 
f_j = A_j^{\a_j /2} T_{\a_j}(B_j / A_j^{1/2}),\qquad g_j = A_j^{(\a_j-1) /2} (B_j^2-A_j)^{1/2} U_{\a_j-1}(B_j / A_j^{1/2}),
\ee
where $\a_0=0$ and 
\be \ba\label{A_0=}
A_0 &=-64 r \frac{s_{13}s_{14}s_{24}s_{25}s_{35}}{s_{12}s_{15}s_{23}s_{34}s_{45}} ,\quad A_i = \frac{s_{i,i+2}s_{i+1,i+4}}{s_{i,i+1}s_{i+2,i+4}},\\
B_0 &= \sum_{i=1}^5 \frac{4}{s_{i,i+1} s_{i+2,i+3}}, \quad 
2B_i = \frac{s_{i+1,i+3}s_{i+2,i+3}}{s_{i,i+1,i+2,i+4}}
-\frac{s_{i,i+2}}{s_{i+2,i+4}}
-\frac{s_{i+1,i+4}}{s_{i,i+1}},~i=1,\ldots, 5.
\ea \ee
In \eqref{A_0=}, the factor $r=s_{12}^2 (s_{13} - s_{24})^2 + (s_{13} s_{25} + s_{15} (s_{24} + s_{25}))^2 + 
 2 s_{12} (s_{13}^2 s_{25} + s_{15} s_{24} (s_{24} + s_{25}) - 
    s_{13} (s_{15} s_{24} - s_{15} s_{25} + s_{24} s_{25}))$ is associated to the sum of the roots of the determinant in the denominator of \eqref{fundamental}.

The proof of \eqref{fundamental2} is a straightforward application of the expressions of the Chebyshev polynomials
\be\ba 
T_n(x) &= \frac{(x-\sqrt{x^2-1})^n+(x+\sqrt{x^2-1})^n}{2},\\
U_n(x) &= \frac{(x+\sqrt{x^2-1})^{n+1}-(x-\sqrt{x^2-1})^{n+1}}{2},\\
\ea\ee
and the fact that for $n=2$ the scattering equations are quadratic. 

The expression \eqref{fundamental2} opens up the possibility of recursion relations among different $\a_i$s due to the recursion relations that the Chebyshev polynomials satisfy.  We will not pursue this further here.

\section{Evaluation of contour integrals at special kinematics}
In this section we will choose a special kinematics that will enable the evaluation of any contour integral on ${\cal M}_{0,n}$ via linear operations without the need to explicitly solve the scattering equations.  In order to achieve this we demand that the polynomial form of the scattering equations becomes a linear system of all symmetric polynomials formed by the variables of the problem. 

We start with the polynomial form of the scattering equations \eqref{scattering eqns} 
\be\label{polynomial form}
\sum_{S \subset A,~|S|=m} k_S^2 \s_S = 0, \qquad 2 \leq m \leq n-2,
\ee
where $A={1,2,\ldots,n}$ and the sum runs over all subsets of $S$ with $m$ elements.  Furthermore
\be
k_S = \sum_{a\in S} k_a,\qquad \s_S = \prod_{b\in S}\s_b.
\ee
There are $n!/(m!(n-m)!)$ elements in each of the $n-3$ equations in \eqref{polynomial form}.  We choose to fix the $SL(2,\mathbb{C})$ invariance by specifying arbitrary values to $\s_1,\s_2,\s_3$.  We choose our special kinematics by demanding that all the $\s$ dependence in \eqref{polynomial form} be written as symmetric polynomials.  We explicitly demonstrate this process for the first few cases.

For the first non-trivial case of $n=5$ we have from \eqref{polynomial form} the following two scattering equations
\be\ba \label{polynomial form n=5}
s_{12} \s_1 \s_2 + s_{13} \s_1 \s_3 + s_{14} \s_1 \s_4 + s_{15} \s_1 \s_5 + s_{23} \s_2 \s_3 + s_{24} \s_2 \s_4& \\
 +s_{25} \s_2 \s_5 + s_{34} \s_3 \s_4 + s_{35} \s_3 \s_5 + s_{45} \s_4 \s_5&=0, \\
s_{123}\s_1\s_2\s_3 + s_{124}\s_1\s_2\s_4 + s_{125}\s_1\s_2\s_5 + s_{134}\s_1\s_3\s_4 +s_{135}\s_1\s_3\s_5 +s_{145}\s_1\s_4\s_5& \\
+s_{234}\s_2\s_3\s_4 + s_{235}\s_2\s_3\s_5 + s_{245}\s_2\s_4\s_5 + s_{345}\s_3\s_4\s_5 &=0,
 \ea\ee
where we have used the notation $s_{ij\ldots}=(k_1+k_2+\ldots)^2$.  By equating the coefficients of $\s_4$ and $\s_5$ in \eqref{polynomial form n=5} we get the following system of equations for the various kinematic invariants
\be\ba 
s_{14}\s_1+s_{24}\s_2 & = s_{15}\s_1 + s_{25}\s_2, \\
s_{124}\s_1\s_2 +s_{134}\s_1\s_3+s_{234} \s_2\s_3 &=s_{125}\s_1\s_5+s_{135}\s_1\s_3+s_{235}\s_2\s_3,
\ea\ee
which has the solution
\be 
s_{14}=s_{15},\qquad s_{24}=s_{25}.
\ee
For that special kinematics it is easy to see that the scattering equations take the form
\be 
A_{2\times 2} 
\begin{pmatrix}
s_1 \\ s_2
\end{pmatrix}
= B_{2\times 1} ,
\ee
where the matrices $A,~B$ depend on kinematics and gauge fixing and $s_i$ denote symmetric polynomials, $s_1 = \s_4+\s_5+\s_6+\ldots$, $s_2 = \s_4 \s_5 + \s_4 \s_6+\s_5\s_6+\ldots$, etc..  The exact expression of $A$ and $B$ is not important to write it down explicitly.

For the next case, $n=6$, we have three scattering equations
\be\ba 
s_{12} \s_1 \s_2 + \ldots = s_{123}\s_1\s_2\s_3 + \ldots = 
s_{1234}\s_1\s_2\s_3\s_4 + \ldots = 0.
\ea\ee
For the choice of special kinematics
\be 
s_{14}=s_{15}=s_{16},\quad s_{24}=s_{25}=s_{26},\quad s_{45}=s_{46}=s_{56}
\ee
the scattering equations take the linear in symmetric polynomials form
\be \label{A3=}
A_{3\times 3} 
\begin{pmatrix}
s_1 \\ s_2 \\s_3
\end{pmatrix}
= B_{3\times 1}.
\ee

Similarly, for $n=7$ we get the following special kinematics
\be 
s_{14}=s_{15}=s_{16}=s_{17},\quad s_{24}=s_{25}=s_{26}=s_{27},\quad s_{45}=s_{46}=s_{47}=s_{56}=s_{57}=s_{67}.
\ee
The general case is straightforward.  Without loss of generality we choose to set $s_{ij}=1,~i,j\geq 4$.  Then, using conservation of momentum and on-shell conditions we can determine all kinematic invariants.  They depend on two parameters and we explicitly present them in the following table
\begin{table}[H]
\centering
\begin{tabular}{l r}\hline\hline
 $s_{12} = (3-n)(\a+\b+n-2)/2$  \\
 $s_{13} = (n-3)(n-3+\a)/2$   \\
 $s_{23} = (n-3)(n-3+\b)/2$  \\
 $s_{1a} = (1+\b)/2$ & $a\geq 4$\\
 $s_{2a} = (1+\a)/2$ & $a\geq 4$\\
 $s_{3a} = (6-2n-\a-\b)/2$ & $a\geq 4$\\
 $s_{ab} = 1$ & $a,b\geq 4, a\neq b $\\
 \hline
\end{tabular}
\caption{Our two parameter special kinematics.}
\label{special_kinematics}
\end{table}

In the above $\a$ and $\b$ are arbitrary real parameters.  We observe that the above special kinematics is exactly the one considered in \cite{Kalousios:2013eca}.  There, it was imposed that $\a,\b>-1$, but here we do not have to do this.  Consequently, the formulas presented in \cite{Kalousios:2013eca} should hold for any value of $\a,\b$.  Another difference with \cite{Kalousios:2013eca} is the fact that there a special gauge fixing was used, namely $\s_1=-1,~\s_2=1,\s_3=\infty$, but here we did not have to use any specific gauge fixing.

We now proceed to the evaluation of the contour integrals.  We have argued before that in our special kinematics the scattering equations are linear in the symmetric polynomials formed from the variables of the scattering equations.  Assume that $(\ldots,r_i,\ldots,r_j,\ldots)$ is a solution of the scattering equations.  Then, due to the symmetry of the problem, so does $(\ldots,r_j,\ldots,r_i,\ldots)$.  This means that if we know one of the solutions of the scattering equations, the rest can be formed from all possible permutations of that solutions.  The number of permutations is $(n-3)!$ which is also the number of all solutions of the scattering equations \cite{Cachazo:2013iaa,Cachazo:2013gna}.  Furthermore, any contour integral $I$ has the following general form
\be\ba\label{I=}
I &= \sum_{\rm solutions} f(\s_4,\s_5,\s_6,\ldots) \\
&= \sum_{\s_i \rm{~perms}} f(\s_4,\s_5,\s_5,\ldots) \\
&= F(s_1,s_2,s_3,\ldots),
\ea\ee
where $f$ is any arbitrary function consistent with $SL(2, \mathbb{C})$ invariance and $F$ a function that depends on symmetric polynomials.  The arguments in $F$ start from $\s_4$ due to our gauge fixing.  The last equality of \eqref{I=} comes from the fundamental theorem of symmetric polynomials.  Therefore, we have proven that any given contour integral can be expressed as a function of symmetric polynomials.  We have seen earlier that the scattering equations is a linear system of the symmetric polynomials, which can be easily computed.  Thus, we can explicitly evaluate any contour integral.  The above discussion can be also used to prove the formulas in \cite{Kalousios:2013eca}, which were first found numerically.

Let us demonstrate the above technique with two examples.  We first want to evaluate the quantity
\be 
I_1 = \sum_{{\mathrm{solutions}}} \frac{\s_{14}\s_{56}}{\s_{15}\s_{46}}.
\ee
According to our previous discussion we have to sum over all six permutations of indices $\s_4,\s_5,\s_6$.  Then we get immediately the desired answer to be $I_1=3$.  We now chose a more complicated example
\be 
I_2 = \sum_{{\mathrm{solutions}}} \frac{\s_{12}\s_{34}}{\s_{13}\s_{24}} = \sum_{{\mathrm{perms}}} \frac{\s_{12}\s_{34}}{\s_{13}\s_{24}}.
\ee
According to the fundamental theorem of symmetric polynomials $I_2$ can be expressed as a function of symmetric polynomials.  We find
\be 
I_2 = 
\frac{2\s_{12}}{\s_{13}}
\frac{
\s_2(\s_2+2\s_3)s_1-(2\s_2+\s_3)s_2+3 s_3-3\s_2^2\s_3
}
{
\s_2^2 s_1-\s_2 s_2+s_3-\s_2^2
}.
\ee 
Substitution of the $s_1,s_2,s_3$ from \eqref{A3=} yields the final answer
\be 
I_2 = \frac{6(4+\a+\b)}{1+\a}.
\ee
Of course, first fixing the gauge and then performing the computation can be more efficient and one can pursue this path in his calculations.

\section{Comments on the companion matrix method}
In this section we will reconstruct the companion matrices discussed in \cite{Huang:2015yka} by using the elimination theory.  We will then move on to establish the equivalence of the method with the algorithm presented in {\cite{Kalousios:2015fya}.

\subsection{Systematic construction of the companion matrices from the elimination theory}\label{CCM}
We will show how to systematically construct the companion matrices of \cite{Huang:2015yka} in an alternative way. We start with some definitions (the reader may refer to \cite{Sturmfels:2002,Froberg:1998} for more details).

Primary definitions: Let $\{\s_1,\s_2,\cdots\,\s_n\}$ be a set of arbitrary $n$ points on the complex plane. Due to $SL(2, \mathbb{C})$ invariance of the massless scattering amplitudes we can fix three of the $\s$s to arbitrary values that we choose to be $\s_1=\infty,\s_2=1,\s_{n}=0$. Let $S:={\cal Q}_{N}[\sigma_3,\cdots,\sigma_{n-1}]$ be the set of polynomials of order $N$ in the variables $\vec{\sigma}:=(\sigma_3,\cdots,\sigma_{n-1})$ with coefficients in ${\cal Q}$ (for our purposes ${\cal Q}=\mathbb{C})$   and let $I$ be an ideal in $S$ whose  Gr\"{o}bner basis is denoted by $G$. We can associate a monomial basis $B$ for the space $S_I:=S/I$. Thinking of $S_I$ as a vectorial space of dimension $d$, $B$ is a vector basis, i.e, any polynomial in $S_I$ can be written in terms of the monomial components of $B$.

The linear transformation (endomorphism) 
\be (f\in S/I)\,\to\,\sigma_i\, f\quad\,\,i=3,\ldots,n-1
\ee can be represented in terms of the basis $B$ by a $d\times d$ matrix $T_{i}$. Specifically, the $l^{\rm th}$ row of the matrix $T_i$ is given by the components in basis $B=(B_1,\ldots,B_{(n-3)!})$ of the polynomial reduction of $\sigma_i\, B_{l}$ with respect to $G$. The matrices $T_i$ are known as the companion matrices. 

In the case of the scattering equations \eqref{scattering eqns} the order of the space is $(n-3)!$ and the ideal in $S$ is given by the scattering equations $I=\{f_a,~ a=3,\ldots,n-1\}$. We can use the polynomial form of the scattering equations \eqref{polynomial form} and then use the elimination theory \cite{Dolan:2014ega} to get an equivalent set of polynomials composed by a one variable $(n-3)!$ degree polynomial in namely $\s_{n-1}$, and $(n-4)$ polynomials in the two-variables $(\s_{n-1},\,\s_{j})$, each of them linear in the remaining variables $\s_j,~j=3,\ldots,(n-2)$. This can also be equivalently done by constructing the associated  Gr\"{o}bner basis (for example by means of the Buchberger method \cite{Buchberger:1985}). This process allows us to write all the solutions for the variables $\s_j$ as functions of the $\s_{n-1}$ variable. This in turn implies that the monomial basis $B$ for the space $S_I$, can be completely expressed in terms of one single variable $\s_{n-1}$ as
\be
B=(1,\s_{n-1},\s_{n-1}^2,\ldots,\s_{n-1}^{(n-3)!})\,.
\ee

The $l^{\rm th}$ row of the companion matrix $T_{\sigma_{n-1}}$ is the vector made out of the components, with respect to the basis $B$, of the polynomial reduction of the monomial $\vec{v}_{n-1}\,B[l]=\s_{n-1}\,B[l]$ with respect to the Gr\"{o}bner basis of $S$. The reduction process consists of lowering the degree\footnote{In this note we use the usual lexicographical order.} of the given monomial by using the polynomials in the Gr\"{o}bner basis.  The only possible polynomial adequate to decrease the degree of $\s_{n-1}\,B[i]$ is the one variable $(n-3)!$ degree polynomial in the Gr\"{o}bner basis, otherwise the reduction can not be written as an expansion in $B$ because reducing with other polynomial will mix the variable $\s_{n-1}$ with the rest (and hence, increasing its degree when rewriting the mixing in terms of $\s_{n-1}$).  
All components of the vector $B$ are of degree equal or lower than $(n-3)!$, which means that the  reduction of the monomials $\s_{n-1} B$ with respect to $G$ is the monomial itself except for the highest component of $B$.  For the highest component $B_{(n-3)!}$ of order $(n-3)!$ one can reduce it by one order by solving the degree $(n-3)!$ single variable polynomial obtained from elimination.

The whole analysis above implies that the form of the companion matrices corresponding to the ``special" variable $\s_{n-1}$ is uniquely fixed by the one variable $(n-3)!$ degree polynomial obtained from the elimination process or from the Gr\"{o}bner basis. Let that polynomial be given by the expression $\a_0+\a_1\s_{n-1}+\a_2\s_{n-1}^2+\cdots+\a_{(n-3)!}\s_{n-1}^{(n-3)!}$. Then, the corresponding companion matrix has the following form,
\beqs\label{Tn=}
T_{\sigma_{n-1}}=\left( \begin{array}{cccccc}
0 & 1 & 0 & 0 &\cdots & 0\\
0 & 0 & 1 & 0 &\cdots & 0\\
\cdots &\cdots &\cdots &\cdots &\cdots &\cdots\\
-\frac{\a_0}{\a_{(n-3)!}} & -\frac{\a_1}{\a_{(n-3)!}} & -\frac{\a_2}{\a_{(n-3)!}}&\cdots&\cdots&-\frac{\a_{[(n-3)!-1]}}{\a_{(n-3)!}} \end{array} \right).
\eeqs
We now recall the important theorem by Stickelberger in the form stated in \cite{Sturmfels:2002} and pointed out by \cite{Huang:2015yka} in the context of the scattering equations,\newpage
{\bf Stickelberger's Theorem}: {\it The complex zeroes of the ideal $I$ are the eigenvalues of the companion matrices.}\newline
In other words the matrices $T_{\sigma_i}$ can be rewritten as 
\be\label{Stick} 
T_{\sigma_i}=\rm{Diag}({}_1\sigma_i,\, {}_2\sigma_i,\,\cdots,\,{}_{(n-3)!}\sigma_i)\,, 
\ee
where ${}_j\sigma_i$ denotes the $j^{\rm th}$ root in the variable $\sigma_i$. This shows that the companion matrices are a way to pack the roots into a nicely organized way.  From (\ref{Stick}) it is obvious that the companion matrices obey any equation satisfied by the roots themselves. 

The last observation allows us to built all the companion matrices in the following way.  Once one get the matrix $T_{\sigma_{n-1}}$ by the method described above, we replace the variable $\s_{n-1}$ for this matrix in the set of polynomials obtained from the elimination method. After the replacement we can use those polynomials equations to solve for the rest of the variables in terms of the matrix $T_{\sigma_{n-1}}$ and that matrix solution should correspond to the companion matrices of the respective variables.

Let us now look at how to apply this to the case of the scattering amplitudes.  We know that $n$-point massless scattering can be expressed in the general form
\be 
{\cal A}_n=\sum_{\sigma_{sol}}{\cal F}(s_{ij},\s_{ij},\epsilon)\,.
\ee
This expression can always be rewritten as a rational function of polynomials as
\be 
{\cal A}_n=\sum_{\s_{sol}}\frac{P({}_{j_1}\s_3,\,{}_{j_2}\s_4,\cdots,\,{}_{j_{(n-3)!}}\s_{(n-3)!})}{Q({}_{j_1}\s_3,\,{}_{j_2}\s_4,\cdots,\,{}_{j_{(n-3)!}}\s_{(n-3)!})}\,,
\ee
where we should replace the solutions ${}_j\s_i$ in terms of their corresponding companion matrices and the factor $Q^{-1}$ should be understood as an inverse matrix. Recalling that all the companion matrices can be expressed in terms of $T_{\sigma_{n-1}}$ we can rewrite the amplitude schematically as,
\be\label{CMA} {\cal A}_n={\rm Tr}\left(\tilde{P}(T_{\sigma_{n-1}})\tilde{Q}^{-1}(T_{\sigma_{n-1}})\right)\equiv{\rm Tr}\left({\cal P}(T_{\sigma_{n-1}})\right)\,.\ee
In the last equation we have defined ${\cal P}$ as the polynomial resulting from the product $\tilde{P}\,\tilde{Q}^{-1}$, emphasizing that the amplitude is reduced to a trace over a single matrix.

For the sake of completeness we display here the first non-trivial example of $n=5$. After using momentum conservation, the polynomial form of the gauge fixed scattering equations are given by,
\be\ba\label{linear5}
h_1&=s_{12}+s_{13}\s_3+s_{14}\s_4\,,\\
h_2&=s_{45}\s_3+s_{35}\s_4+s_{25}\s_3\s_4\,.
\ea\ee
Elimination theory implies that the single variable polynomial is given by
\be 
\left| \begin{array}{cc}
 h_1&h_2\\
 \partial_{\s_3}h_1&\partial_{\s_3}h_2\end{array} \right|=\s_4^2\,s_{14}s_{25}+\s_4\,(s_{12}s_{25}-s_{13}s_{35}+s_{14}s_{45})+s_{12}s_{45}=0.
\ee
Putting the coefficients of this polynomial into the matrix (\ref{Tn=}) we have
\be 
T_4=\left( \begin{array}{cc}
 0&1\\[-0.4cm]\\
-\frac{s_{12}s_{45}}{s_{14}s_{25}}&-\frac{s_{12}s_{25}-s_{13}s_{35}+s_{14}s_{45}}{s_{14}s_{25}}\end{array} \right).
\ee
Solving the first equation in (\ref{linear5}) for $\s_3$ we get
\be \s_3=-\frac{s_{14}}{s_{13}}\s_4-\frac{s_{12}}{s_{13}},\ee
and replacing $\s_4$ by $T_4$ in the last equation give us,
\be 
T_3=\left( \begin{array}{cc}
 -\frac{s_{12}}{s_{13}}&-\frac{s_{14}}{s_{13}}\\[-0.15cm]\\
\frac{s_{12} s_{45}}{s_{13} s_{25}}&\frac{-s_{13} s_{35} + s_{14} s_{45}}{s_{13} s_{25}}\end{array} \right), \ee
which obviously coincides with the results in \cite{Huang:2015yka}.  The next case of $n=6$ is equally straightforward, but we will not work it out here.

\subsection{Equivalence between the companion matrix method and the algorithm of \cite{Kalousios:2015fya}}
Having discussed the relation between the elimination theory and the companion matrix method, we would now like to discuss the equivalence of the method to the algorithm presented in \cite{Kalousios:2015fya}, which was based on the idea that one can express any amplitude in terms of a particular combination of coefficients of polynomials given essentially by the Vieta formulas.  The procedure of \cite{Kalousios:2015fya} allowed the construction of any amplitude as a rational function of the kinematic invariants without the need to solve the scattering equations.  This is also what the companion matrix achieves. Following subsection \ref{CCM} above, the amplitude has been reduce to a trace over a single matrix corresponding to one of the variables $\s_{n-1}$ (\ref{CMA}). Let us first go through some simple examples to see the equivalence between the two methods. The simplest case we can consider is 
\be 
\sum_{\rm roots} \s_{n-1} = {\rm Tr}\, (T_{n-1}) = -\frac{\a_{[(n-3)!-1]}}{\a_{(n-3)!}}
\ee
as it follows from \eqref{Tn=}.  This is simply the coefficient of the $\s_{n-1}$ polynomial as discussed in \cite{Kalousios:2015fya}.  Moving to the next example we will consider
\be 
\sum_{\rm roots} \s_{n-1}^2 = {\rm Tr}\, (T_{n-1}^2).
\ee
Here taking the square of the companion matrix mixes the different coefficients of the  $\s_{n-1}$ polynomial in a consistent way giving finally the answer
\be\ba
\sum_{\rm roots} \s_{n-1}^2 &= {}_1\s_{n-1}^2+{}_2\s_{n-1}^2+{}_3\s_{n-1}^2+\ldots = ({}_1\s_{n-1}+{}_2\s_{n-1}+{}_3\s_{n-1}+\ldots)^2\\
&\qquad -2({}_1\s_{n-1}\, {}_2\s_{n-1}+{}_1\s_{n-1}\, {}_3\s_{n-1}+{}_2\s_{n-1}\, {}_3\s_{n-1}+\ldots),
\ea\ee
with ${}_i\s_{n-1}$ denoting the roots of the $\s_{n-1}$ polynomial and the answer follows from the Vieta formulas \cite{Kalousios:2015fya}.  An overall sign due to different conventions is immaterial.

More complicated examples follow straightforwardly.  This shows that due to the form of the $T_{n-1}$ companion matrix taking the trace is an organized way to realize the algorithm of \cite{Kalousios:2015fya}.

We make one more comparison.  In \cite{Kalousios:2015fya} it was shown that a way to organize the amplitudes is to consider a fundamental quantity that depended on cross ratios raised to an arbitrary power.  The computation and organization of the various terms was shown to be captured via a generating function.  Derivatives of the generating function there correspond to taking powers of companion matrices here.

\vspace{0.5cm}
{\it Sketch of a general proof:} In equation (\ref{CMA}) we showed that the amplitude is reduced to a trace over one matrix which is computed as a polynomial of a single building block matrix that we will denote simply as $T$ and which has the general form (\ref{Tn=}). In general, we can rewrite equation (\ref{CMA}) as
\be\label{CMAE} {\cal A}_n={\rm Tr}\left({\cal P}(T_{\sigma_{n-1}})\right)=A_0+A_1{\rm Tr}(T)+A_2{\rm Tr}(T^2)+\cdots+A_{q}{\rm Tr}(T^q)\,,\ee

where $q$ is an integer\footnote{The integer $q$ depends on $n$, but we do not explicitly show this dependence because it is not important for the purpose of this section.} and we have used the property of the trace
\be {\rm Tr}(\sum_i T_i)=\sum_i {\rm Tr}(T_i) \,.\ee

From (\ref{CMAE}) we see that it is enough to prove the equivalence by considering an arbitrary power of $T$.
This is somehow trivial after using Stickelberger's theorem on the single variable polynomial. From the theorem we know the eigenvalues of the matrix $T$ will be given by the roots of the polynomial, which implies 
\be\label{equiv} {\rm Tr}(T_{\s_i}^q)=\sum_j({}_j\s_i)^q
=\left(\sum_j{}_j\s_i\right)^q-\hspace{-0.5cm}
\sum_{q_1+q_2+\cdots+q_n=q}^{q_i\neq n}\frac{q!}{q_1!q_2!\cdots q_n!} \prod_{j}{}_j\s_i^{q_j},
\ee
with $ q_i\in\mathbb{N}$.  The last line in (\ref{equiv}) can immediately be 
rewritten in terms of the coefficients by using the Vieta formulas on the right hand side.  Ratios of $\s$s follow similarly.

\section{Conclusions}
In this note we discussed three ideas.  The first one was the interesting observation that the most fundamental quantity for five point amplitudes consistent with gauge invariance can be expressed in terms of Chebyshev polynomials of the first and second kind.  This followed from a straightforward computation and properties of the polynomials.  Furthermore and since the polynomials satisfy recursion relations, this opens up the possibility or recursion relation among expressions raised to different powers of the cross ratios.

The second idea was the choice of a special kinematics that transformed the polynomial form of the scattering equations to a linear system of symmetric polynomials.  Then due to the symmetry of the problem, knowledge of one solution of the scattering equations is enough to determine all of the solutions through the $(n-3)!$ possible permutations of that one solution.  We then argued that any amplitude can be written as a function of symmetric polynomials only, which eventually makes it possible its calculation through linear operations.  The special kinematics turned out to be the one previously discussed in \cite{Kalousios:2013eca} with respect to roots of Jacobi polynomials.  Here we were able to generalize the formulas presented there and we set a framework for a possible proof of relations discussed there that were found numerically.

We did not discuss in this work the possibility of using symmetric polynomials for general kinematics but it is definitely an interesting idea to consider.  It seems that for the first non trivial case of $n=5$ it is easy to find a change of variables that transforms the scattering equations to a linear system of symmetric polynomials, but it remains to be seen if this can be generalized to general $n$.

The last idea we discussed was related to the recently presented method of companion matrices {\cite{Huang:2015yka}.  We argued that the last companion matrix consists only of the coefficients of the $(n-3)!$ order one variable polynomial, whereas zeroes and ones fill in the rest of the matrix.  All of the companion matrices can be produced using either the elimination theory presented in \cite{Dolan:2014ega} or the Gr\"{o}bner basis presented in \cite{Huang:2015yka}.  Furthermore we argued that due to the form of the last companion matrix, the method presented in \cite{Huang:2015yka} is essentially equivalent to reading coefficients of polynomials as it was first presented in \cite{Kalousios:2015fya}.  Definitely, the companion matrix method is a systematic way to realize the elimination theory and the algorithm of \cite{Kalousios:2015fya}, but it is restricted by computer power.  Since using the elimination theory or calculating the Gr\"{o}bner basis is typically a very time consuming process for large polynomial systems, it remains to be seen if there is an efficient way to construct the appropriate polynomials needed for the evaluation of the amplitudes.

\vspace{5mm}

\noindent
{\bf Acknowledgments}

\vspace{3mm}

\noindent
It is a pleasure to thank Humberto Gomez, Wei He and Francisco Rojas for useful discussions.  The work of C.C. is supported in part by the National Center for Theoretical Science (NCTS), Taiwan.  The work of C.K. is supported by the S\~ao Paulo Research Foundation (FAPESP) under grants 2011/11973-4 and 2012/00756-5.

\bibliographystyle{utphys}
\bibliography{mybib}
\end{document}